\documentclass[trackchanges, twocolumn, twocolappendix]{aastex701}

\usepackage{newtxtext,newtxmath}
\usepackage[T1]{fontenc}
\usepackage{amsmath,booktabs,graphicx,tabularx}
\hypersetup{colorlinks=true,citecolor=blue,linkcolor=blue,urlcolor=blue}

\newcommand{\dd}{\mathrm{d}}

\makeatletter
\protected\def\letterstrikeout{\leavevmode\bgroup\ULdepth=-.55ex\ULset}
\makeatother

\newcommand{\VMinimalTauLowerDays}{524}

\begin{document}

\title{Testing internal dynamics in neutron stars with radio pulsar timing}

\author[orcid=0000-0001-8589-6851,sname='Reginald Christian Bernardo']{Reginald Christian Bernardo}
\affiliation{Max Planck Institute for Gravitational Physics (Albert Einstein Institute), \\ Leibniz Universit\"{a}t Hannover, Callinstrasse 38, D-30167 Hannover, Germany}
\email[show]{reginald.christian.bernardo@aei.mpg.de}


\begin{abstract}
Rotational variations in pulsars reflect either internal angular momentum exchange via equal and opposite torques (stationary spin noise) or independent torques causing diffusive changes in total angular momentum (nonstationary spin noise). We compare two models of a radio-emitting crust and interior superfluid to test the stationarity of angular velocity fluctuations through observed pulsar phases. We present results for three isolated pulsars with measurable timing noise, demonstrating distinct inference outcomes. For PSR J1359-6038, the data mildly favor independent torques over equal-and-opposite torques, with an odds ratio of seven hundred to one. For PSR B1937+21, the analysis shows no clear preference between the models. For Vela in a pre-glitch interval, excess scatter and unexplained residual structure limit the interpretation.
\end{abstract}


\section{Introduction}
\label{sec:introduction}

Pulsar timing measurements reveal stochastic rotational variations that reflect torques acting on neutron stars \citep{1975ApJS...29..453G,2010MNRAS.402.1027H}. Radio pulses, emitted from the magnetosphere, track the rotation of the charged crust. However, neutron stars contain a superfluid interior that can rotate differentially and exchange angular momentum with the crust through mutual friction \citep{Prix:1999gk,Prix:2002jn,Andersson:2005rs,Haskell:2017lkl}. Thus, observed crustal fluctuations may arise from internal angular momentum redistribution within the star or from external torques that alter its total angular momentum.

We compare two dynamical two-component models of pulsar rotation that differ in their stochastic torque assumptions. In the singular model, independent torques act on the crust and superfluid, driving a diffusive total angular momentum mode and producing nonstationary spin noise, along with a damped crust-core spin lag \citep{Meyers:2021myb,Meyers:2021slh,Bernardo:2026opv}. In the minimal model, equal and opposite torques conserve total angular momentum, producing stationary spin noise \citep{Antonelli:2022gqw,Antonelli:2025vhw}. Both models include mutual friction between the crust and superfluid, enabling angular momentum exchange. The key difference lies in the torque correlations, which determine the stationarity of the spin noise: nonstationary for the singular model, stationary for the minimal model. Our inference tests these torque statistics and their associated rotational dynamics.

Correlations between radio pulsar timing and continuous gravitational wave signals have been developed as probes of neutron star interiors \citep{Haskell:2017lkl,2021arXiv210403137H,Antonelli:2025vhw}. We infer the torque models through radio timing and reconstruct their predictions for the unobserved component. We apply the framework to three isolated pulsars, chosen to illustrate different inference outcomes and to represent diverse astrophysical contexts: J1359-6038, a reference timing-noise case \citep{ONeill:2024uiw}; B1937+21, a millisecond pulsar central to nanohertz gravitational wave searches \citep{NANOGrav:2023hde}; and Vela, a frequent glitcher whose glitches provide information about superfluid reservoirs \citep{1969Natur.224..872B,2012PhRvL.109x1103A,2013PhRvL.110a1101C,2020A&A...642A.223M}.

Table~\ref{tab:headline} presents the principal results. Priors and detailed constraints (marginalized statistics) appear in Table~\ref{tab:phase-results} in Appendix~\ref{sec:priors}. Appendix~\ref{sec:methods} describes the spin noise models, likelihood and inference. Appendix \ref{sec:power-spectra} describes the power spectra that can be associated with the two-component models.

\section{Data}
\label{sec:data}

PSR J1359-6038, B1937+21 and Vela are isolated radio pulsars.
J1359-6038 has a rotation period of about 127 ms and characteristic age of 319 kyr; it provides a published case of measurable timing noise without a glitch in this interval \citep{ONeill:2024uiw,Dong:2025nho}. B1937+21 rotates in about 1.56 ms and extends the comparison to an isolated millisecond pulsar \citep{NANOGrav:2023hde}.
Vela is younger, with period 89 ms and characteristic age 11.6 kyr.

We use 429 phase measurements of J1359-6038 over 1263 days \citep{2020MNRAS.494..228L} and 382 measurements of Vela over 307 days before the glitch at MJD 57734.48 \citep{2025PASA...42..169T, Sarkissian2017Vela}.\footnote{
\href{https://github.com/Molonglo/TimingDataRelease1/}{https://github.com/Molonglo/TimingDataRelease1/} \\
\href{https://doi.org/10.4225/08/59183E949E033}{https://doi.org/10.4225/08/59183E949E033}
} Their mean consecutive separations are 2.95 and 0.805 days.
The median uncertainties in the times of arrival are 47.6 and 6.29 microseconds, respectively. For these two data sets, we use \texttt{TEMPO2} \citep{Hobbs:2006cd,Edwards:2006zg} with the corresponding timing solutions to obtain residuals, which we express as rotational phases.

For B1937+21, we use the NANOGrav 15 yr HDF5 timing product with 23,023 times of arrival \citep{NANOGrav:2023hde}, retaining its residuals and timing design. Measurements separated by gaps of at most five minutes are grouped into 623 observing epochs spanning 15.87 years. After applying the released noise calibration and combining the TOAs within each epoch, the median uncertainty of the epoch averaged measurements is $0.117\,\mu{\rm s}$. The released residuals include the piecewise dispersion correction, DMX. Appendix~\ref{sec:methods} describes the treatment of DMX, phase conversion, noise calibration and averaging.

\section{Results}
\label{sec:results}

\begin{table}
\centering
\caption{
  Log Bayes factors, $\ln B_{s,m} = \ln(Z_s/Z_m)$, comparing singular (nonstationary spin noise) and minimal (stationary spin noise) dynamics.
  Positive values favor singular dynamics; negative values favor minimal dynamics.
  Uncertainties are Monte Carlo standard errors.
}
\label{tab:headline}
\begin{tabular}{lcc}
\toprule
Pulsar & $\ln B_{s,m}$ & Interpretation \\
\midrule
J1359$-$6038 & $6.59 \pm 0.09$ & Mild preference \\
B1937$+$21 & $0.93 \pm 0.04$ & Weak preference \\
Vela, pre-glitch & $17.19 \pm 0.07$ & Noise-limited \\
\bottomrule
\end{tabular}
\end{table}

\subsection{PSR J1359-6038}

\begin{figure}[!htbp]
 \centering
 \includegraphics[width=\columnwidth]{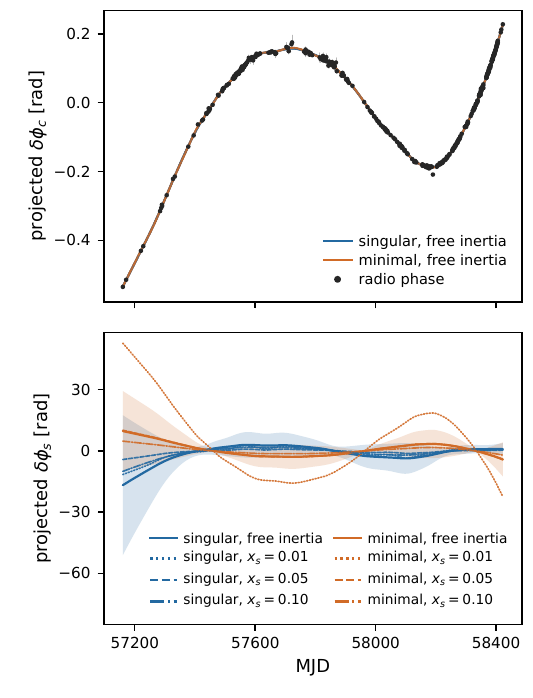}
\caption{
J1359-6038 phase residuals and reconstructions. Solid curves show posterior means and shaded bands one standard deviation. Thin curves show conditional means at fitted parameters for $x_s=0.01$, $0.05$ and $0.10$. Error bars show projected timing uncertainties before the
additional fitted noise rescaling. The lower panel predicts the unobserved superfluid phase.
}
 \label{fig:j1359}
\end{figure}

The radio phases of J1359-6038 mildly favor singular dynamics, with $\ln B_{s,m}=6.59\pm0.09$ (Table~\ref{tab:headline}). This compares the two specified torque models under their adopted priors.

J1359-6038 provides a reference comparison with an earlier Kalman filter based analysis \citep{ONeill:2024uiw}. Utilizing the same priors, our singular model constraints turn out broadly consistent, despite using rotational phases rather than locally estimated angular velocities. We obtain $\log_{10}(\tau/{\rm s})=7.54^{+0.43}_{-0.99}$, corresponding to a median relaxation time of about 400 days and a central 90\% interval of approximately 40--1100 days. The stochastic torque parameters have posterior medians $Q_c\simeq2\times10^{-24}$ and $Q_s\simeq9\times10^{-22}\,{\rm rad^2\,s^{-3}}$, with broad intervals overlapping the Kalman estimates. The ratio of component moments of inertia remains poorly constrained.

Both models follow the radio residuals but predict different superfluid phases (Figure~\ref{fig:j1359}). In minimal dynamics the relative scale is set by the superfluid moment of inertia fraction. Fixing it gives an anticorrelated copy of the crust stochastic phase. Independent torques permit superfluid fluctuations that the crust does not determine. The radio data constrain the two components differently for the two models. The shaded superfluid regions describe predictions conditional on the data and assumed dynamics.

The relaxation time posterior has a long upper tail. Extending this bound from $10^8$ to $10^{10}\,\mathrm{s}$ preserves the preference for singular dynamics, although the numerical Bayes factor decreases. This preference also persists when initialization is moved between 10 and 1000 days before the first observation.

\subsection{PSR B1937+21}
\label{sec:b1937}

\begin{figure}[!htbp]
 \centering
 \includegraphics[width=\columnwidth]{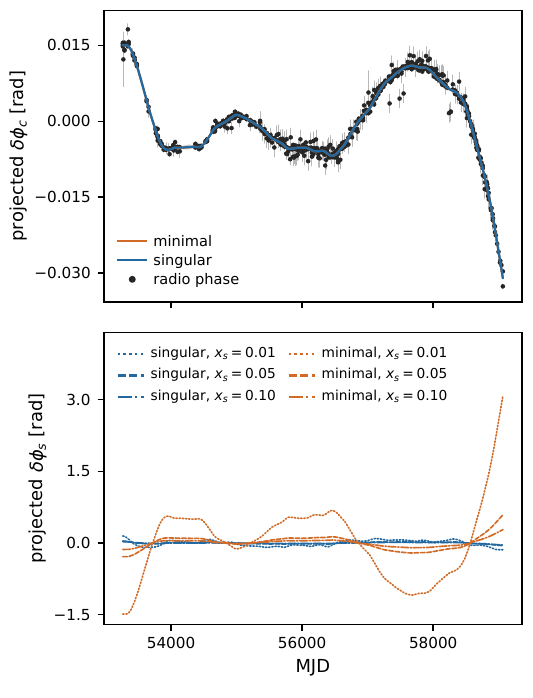}
\caption{
B1937+21 phase reconstructions using the released DMX correction. The upper panel shows posterior means and one standard deviation for both models; the bands are narrow at this scale. Error bars show projected, calibrated timing uncertainties before the
additional fitted noise rescaling. The lower panel shows conditional superfluid means at fitted parameters for $x_s=0.01$, $0.05$ and $0.10$.
}
 \label{fig:b1937}
\end{figure}

For B1937+21, $\ln B_{s,m}=0.93\pm0.04$ gives only a weak, indecisive preference for singular dynamics (Table~\ref{tab:headline}).

The minimal relaxation time has a prior dependent 95 percent lower bound of about 5.5 years and a long upper tail. The combination $\log_{10}[(A_{\min}/\tau)/({\rm rad^2\,s^{-3}})]=-26.72^{+0.15}_{-0.14}$ is more tightly constrained. The median relaxation time is not a resolved coupling measurement. The singular relaxation time and superfluid moment of inertia fraction distributions remain close to their priors.

Allowing the full linearized timing model to vary, including the DMX coefficients, also leaves weak separation. The evidence strength changes when the relaxation time prior is restricted. Additional robustness checks on the timing treatment and prior ranges are reported in Appendices~\ref{sec:priors} and \ref{sec:methods}.

Both crust reconstructions (Figure~\ref{fig:b1937}) have stable posterior means and uncertainty bands. For the singular model, a small number of allowed parameter values produce large superfluid variances that prevented stable marginalized uncertainty band reconstructions for the superfluid phase. Predictions at fixed parameter values remain well defined. The radio data constrain the crust reconstruction. The singular posterior permits parameter values with similar crust predictions but substantially different superfluid fluctuations. The minimal superfluid reconstruction passes the corresponding stability checks. The lower panel shows conditional means at specified superfluid moment of inertia fractions for both models.

\subsection{Vela}

\begin{figure}[!htbp]
 \centering
 \includegraphics[width=\columnwidth]{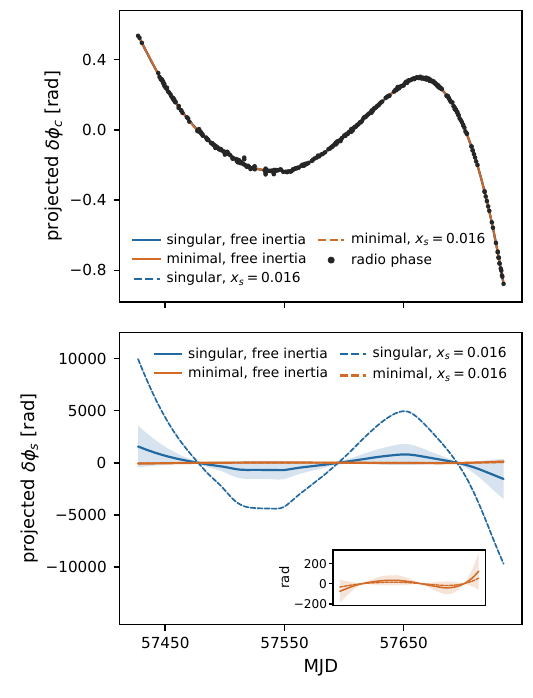}
\caption{
Vela phase reconstructions before the glitch. Solid curves show posterior means and shaded bands one standard deviation. Dashed curves show conditional means at fitted parameters with $x_s=0.016$, a reference motivated by glitch activity under the assumed reservoir model. Error bars show projected timing uncertainties before the
additional fitted noise rescaling. The lower panel predicts the superfluid phase; its inset resolves the minimal model.
}
 \label{fig:vela}
\end{figure}

For Vela, unresolved residual structure limits the physical interpretation of the model comparison, despite a relatively large Bayes factor, $\ln B_{s,m}=17.19\pm0.07$.
However, the posterior median EFAC is approximately 7.6 for singular dynamics and 7.1 for minimal dynamics. This substantial increase in the uncertainty scale accommodates excess scatter, but does not identify its origin. The minimal relaxation time has a prior-dependent 95 percent lower bound of about $\VMinimalTauLowerDays$ days and remains unresolved above this bound.

The large EFAC values and remaining residual structure limit the interpretation of Vela's model preference.
No physical origin for the remaining component is established. Checks of the relaxation time prior and initialization time are reported in Appendices~\ref{sec:priors} and \ref{sec:methods}.

Vela's cumulative glitch activity motivates $x_s=0.016$ as a conditional reference for the superfluid moment of inertia fraction (Figure~\ref{fig:vela}) \citep{2012PhRvL.109x1103A}. Applying this value to the spin noise component assumes that the component responsible for the timing noise participates in the same angular momentum reservoir inferred from glitches. We use $x_s=0.016$ only as a conditional reference. The spin noise models considered neglects entrainment; that modifies the superfluid moment of inertia fraction inferred through glitches \citep{2012PhRvL.109x1103A,2013PhRvL.110a1101C}.

\subsection{Dispersion measure sensitivity}
\label{sec:dm}

The timing data include deterministic dispersion corrections from the published timing models. We additionally test whether stochastic dispersion measure (DM) variations affect the comparison. For J1359-6038 and Vela, previous studies motivate these checks \citep{ONeill:2024uiw,Dong:2025nho,2016MNRAS.459.3104S}, but the limited radio frequency coverage prevents these variations from being constrained independently. We therefore use specified DM amplitudes and temporal spectra. J1359-6038 retains its preference for singular dynamics across the three tested spectra using an amplitude measured from later MeerTime observations \citep{2024MNRAS.530.1581K}. Vela also retains a positive relative Bayes factor under the adopted DM stress test, with a spectrum dependent strength. For B1937+21, the wider frequency coverage allows a DM offset to be fitted within each session without a temporal DM spectrum, giving $\ln B_{s,m}=1.37\pm0.07$. Appendix~\ref{sec:methods} gives the assumed amplitudes and spectra.

\section{Outlook}
\label{sec:falsifiability}

Radio timing probes angular momentum exchange inside a neutron star because the correlations between the stochastic torques determine the fluctuations observed in the crust. Minimal dynamics describe equal and opposite stochastic torques, conserve the total angular momentum and produce stationary angular velocity fluctuations \citep{Antonelli:2022gqw}. Singular dynamics allow independent torques, fluctuations in the total angular momentum and nonstationary angular velocity fluctuations \citep{Meyers:2021myb}. Radio observations, however, directly trace only the crust, and different internal dynamics can produce similar crustal phase histories.

In minimal dynamics \citep{Antonelli:2022gqw}, conservation of total angular momentum fixes $I_c\delta\Omega_c+I_s\delta\Omega_s$ to a constant; or, at the level of the phase
\begin{equation}
 x_c\delta\phi_c+x_s\delta\phi_s=0 \,,
 \label{eq:hidden-phase}
\end{equation}
with $x_c+x_s=1$. A constraint on the participating moment of inertia fraction fixes both the relative amplitude and the opposite sign of the two phase fluctuations. Radio timing alone does not determine this relative scale. An independent measurement of the superfluid phase enables testing the angular momentum exchange hypotheses (spin noise models) directly. An additional constraint on the superfluid moment of inertia fraction would overconstrain the minimal model.

Singular dynamics make a different prediction \citep{Meyers:2021myb}. Even at a fixed superfluid moment of inertia fraction, independent stochastic torques leave a distribution of possible superfluid phase histories. An independent phase measurement could be compared with this distribution and provide information about the relaxation time and stochastic torques that cannot be obtained from the crust alone. The same observation could test either hypothesis: the exact phase relation predicted by minimal dynamics or the conditional phase distribution predicted by singular dynamics.

The three applications separate relative model preference from physical interpretation. J1359-6038 mildly favors singular dynamics, B1937+21 gives weak separation, and Vela remains limited by unresolved noise despite its large relative Bayes factor.
The framework also predicts the rotational behavior of the unobserved component. An independent rotational phase, together with information about the participating superfluid moment of inertia, would extend radio timing into a direct test of internal angular momentum exchange. Continuous gravitational wave observations may provide such a measurement if the emission phase can be tied to the superfluid rotation \citep{2021arXiv210403137H,Antonelli:2025vhw}. Joint radio timing and continuous gravitational wave observations could test whether pulsar spin noise arises from balanced internal torques or independent torques.

\begin{acknowledgments}
The author thanks Boris Goncharov for comments on an earlier version of the manuscript, and Reinhard Prix, Bruce Allen, Rutger van Haasteren and Bjorn Larsen for relevant discussion.
\end{acknowledgments}





\clearpage
\appendix

\section{Priors and inference summaries}
\label{sec:priors}

\begin{table*}
\centering
\caption{
Phase inference for three pulsars.
Entries are posterior medians with central 90\% intervals, or evidences with Monte Carlo standard errors.
Models are singular ($s$) and minimal ($m$); labels $s, x_s^*$ and $m, x_s^*$ condition on the displayed superfluid moment of inertia fraction, while $s$ and $m$ integrate its prior.
Note that the minimal radio likelihood is independent of the superfluid moment of inertia fraction.
The final column gives the posterior-predictive $p$-value ($p_{\rm PPC}$), a goodness-of-fit diagnostic: smaller values indicate poorer agreement between the model and the data (Section \ref{subsec:residual-adequacy}).
Note that $p_{\rm PPC}$ is not the probability that the model is correct.
     }
\label{tab:phase-results}
\setlength{\tabcolsep}{2.6pt}
\renewcommand{\arraystretch}{1.10}
{\begin{tabular}{@{}lcccccccc@{}}

\toprule
Model & $\log_{10}(\tau/{\rm s})$ & $\log_{10}r$ & \shortstack{$s$: $\log_{10}Q_c$\\$m$: $\log_{10}A_{\min}$} & $\log_{10}Q_s$ & $\log_{10}F$ & $\log_{10}(J/{\rm s})$ & $\log Z$ & $p_{\rm PPC}$ \\
\midrule
\multicolumn{9}{l}{\textbf{PSR J1359$-$6038; $x_s^*=0.050$}} \\
\multicolumn{9}{l}{\textit{$s$: priors for singular}: $\log_{10}\tau\in[5.00,8.00]$, $\log_{10}r\in[-2.00,2.00]$, $\log_{10}Q_{c,s}\in[-30.00,-16.00]$} \\
\multicolumn{9}{l}{\textit{$m$: priors for minimal}: $\log_{10}\tau\in[5.00,8.00]$, $\log_{10}A_{\min}\in[-21.70,-11.70]$, $x_s\in[0.001,0.250]$} \\
$s$ & $7.54^{+0.43}_{-0.99}$ & $-0.21^{+1.87}_{-1.64}$ & $-23.72^{+0.28}_{-2.70}$ & $-21.04^{+2.74}_{-1.58}$ & $-0.05^{+0.03}_{-0.03}$ & $-6.35^{+1.29}_{-1.43}$ & $1863.25\pm0.07$  & $0.005$ \\
$s,\ x_s^*$ & $7.59^{+0.37}_{-0.81}$ & -- & $-23.70^{+0.27}_{-0.94}$ & $-19.39^{+0.77}_{-0.92}$ & $-0.05^{+0.03}_{-0.03}$ & $-6.65^{+1.53}_{-1.22}$ & $1863.33\pm0.08$  & $0.006$ \\
$m$ & $7.87^{+0.12}_{-0.28}$ & -- & $-15.40^{+0.21}_{-0.26}$ & -- & $-0.05^{+0.03}_{-0.03}$ & $-6.63^{+1.48}_{-1.27}$ & $1856.66\pm0.06$  & $0.006$ \\
\multicolumn{2}{l}{$\ln B_{s,m}$: free, fixed $x_s$} & \multicolumn{7}{l}{$6.59\pm0.09$\quad  $6.67\pm0.10$} \\
\midrule
\multicolumn{9}{l}{\textbf{PSR B1937$+$21, released DMX; $x_s^*=0.050$}} \\
\multicolumn{9}{l}{\textit{$s$: priors for singular}: $\log_{10}\tau\in[0.00,12.00]$, $\log_{10}r\in[-3.00,-0.33]$, $\log_{10}Q_{c,s}\in[-34.00,-12.00]$} \\
\multicolumn{9}{l}{\textit{$m$: priors for minimal}: $\log_{10}\tau\in[0.00,12.00]$, $\log_{10}A_{\min}\in[-27.00,-3.00]$, $x_s\in[0.001,1/\sqrt{10}]$} \\
$s$ & $5.73^{+5.66}_{-5.01}$ & $-1.77^{+1.24}_{-1.09}$ & $-26.76^{+0.21}_{-5.24}$ & $-25.50^{+5.27}_{-7.42}$ & $-0.05^{+0.02}_{-0.02}$ & $-7.81^{+0.27}_{-0.17}$ & $3633.82\pm0.03$  & $0.006$ \\
$s,\ x_s^*$ & $5.80^{+5.63}_{-5.09}$ & -- & $-26.76^{+0.19}_{-5.17}$ & $-25.62^{+4.63}_{-7.30}$ & $-0.05^{+0.02}_{-0.02}$ & $-7.81^{+0.29}_{-0.17}$ & $3633.74\pm0.07$  & $0.019$ \\
$m$ & $9.94^{+1.71}_{-1.70}$ & -- & $-16.79^{+1.72}_{-1.68}$ & -- & $-0.05^{+0.02}_{-0.02}$ & $-7.80^{+0.28}_{-0.17}$ & $3632.89\pm0.01$  & $0.027$ \\
\multicolumn{2}{l}{$\ln B_{s,m}$: free, fixed $x_s$} & \multicolumn{7}{l}{$0.93\pm0.04$\quad  $0.85\pm0.07$} \\
\midrule
\multicolumn{9}{l}{\textbf{PSR J0835$-$4510 (Vela), before glitch; $x_s^*=0.016$}} \\
\multicolumn{9}{l}{\textit{$s$: priors for singular}: $\log_{10}\tau\in[0.00,12.00]$, $\log_{10}r\in[-3.00,-0.33]$, $\log_{10}Q_{c,s}\in[-34.00,-12.00]$} \\
\multicolumn{9}{l}{\textit{$m$: priors for minimal}: $\log_{10}\tau\in[0.00,12.00]$, $\log_{10}A_{\min}\in[-27.00,-3.00]$, $x_s\in[0.001,1/\sqrt{10}]$} \\
$s$ & $8.26^{+1.32}_{-1.09}$ & $-1.24^{+0.80}_{-1.29}$ & $-22.30^{+0.62}_{-9.28}$ & $-14.14^{+1.77}_{-2.74}$ & $0.88^{+0.03}_{-0.06}$ & $-5.76^{+1.17}_{-1.94}$ & $1470.96\pm0.06$  & $0.006$ \\
$s,\ x_s^*$ & $8.23^{+0.68}_{-1.01}$ & -- & $-22.35^{+0.72}_{-9.89}$ & $-13.65^{+1.35}_{-1.86}$ & $0.88^{+0.03}_{-0.07}$ & $-5.60^{+1.04}_{-2.06}$ & $1471.04\pm0.04$  & $0.006$ \\
$m$ & $9.57^{+1.92}_{-1.91}$ & -- & $-11.32^{+1.94}_{-1.92}$ & -- & $0.85^{+0.03}_{-0.05}$ & $-5.97^{+1.33}_{-1.76}$ & $1453.78\pm0.03$  & $0.005$ \\
\multicolumn{2}{l}{$\ln B_{s,m}$: free, fixed $x_s$} & \multicolumn{7}{l}{$17.19\pm0.07$\quad  $17.26\pm0.05$} \\
\bottomrule
\end{tabular}}
\par\smallskip
\parbox{\textwidth}{\footnotesize
\emph{Notes.}
All displayed logarithmic priors are uniform and normalized; bounds are rounded. Noise priors are $\log_{10}F\in[-1.00,1.00]$ and $\log_{10}(J/{\rm s})\in[-8.00,-2.00]$. Units are $Q_a$ in ${\rm rad^2\,s^{-3}}$ and $A_{\min}$ in ${\rm rad^2\,s^{-2}}$. A dash marks an inapplicable coordinate.
Each $p_{\rm PPC}$ uses 1,024 refitted simulations and has resolution $1/1025$. 
Approximate 95\% Monte Carlo intervals are $[0.001,0.010]$ for entries rounded to 0.005, $[0.002,0.011]$ for 0.006, $[0.010,0.028]$ for 0.019, and $[0.017,0.038]$ for 0.027. The small probabilities have limited numerical precision. 
}
\end{table*}

Table~\ref{tab:phase-results} summarizes the inference results and priors for the minimal (stationary spin noise, internal angular momentum exchange) and singular (nonstationary spin noise, independent torques) models for each pulsar. Uniform priors apply to the logarithmic spin noise coordinates. The minimal radio likelihood is independent of the superfluid moment of inertia fraction, so its normalized prior integrates out and fixed- and free-inertia cases have identical evidences. The hidden phase predictions of both models, however, depend on this prior.

B1937+21 and Vela use broad spin priors. For B1937+21, we also perform conditional fits over a grid of fixed superfluid moment of inertia fractions ($x_{\rm s} = 0.01, 0.02, \cdots 0.10$) using the released DMX product \citep{NANOGrav:2023hde}. The reported conditional comparisons pass their numerical checks (Appendix~\ref{sec:methods}); these fits should not be interpreted as an inertia posterior. Vela is also analyzed with a narrower relaxation time prior.

For B1937+21 we also test how restricting the relaxation time and torque priors affects the evidence. The same numeric bounds are applied to each model's torque coordinates, but because the parameters differ, these are not identical physical priors. We estimate the evidence impact with the fraction of prior and posterior samples inside the restricted region, or the region that was kept after the original prior was trimmed down. Repeating the posterior sampling and evidence calculations with restricted priors turned in consistent results.

\section{Methods}
\label{sec:methods}
We write the spin residuals' (relative to secular rotation) equation of motion as
\begin{align}
 \label{eq:eom_crust} \dd\delta\Omega_c&=-\frac{x_s}{\tau}(\delta\Omega_c-\delta\Omega_s)\dd t+\dd\eta_c \,, \\
 \label{eq:eom_superfluid} \dd\delta\Omega_s&=+\frac{x_c}{\tau}(\delta\Omega_c-\delta\Omega_s)\dd t+\dd\eta_s \,,
\end{align}
where $x_{c,s}=I_{c,s}/(I_c+I_s)$, $r=x_s/x_c$, $\dd\delta\phi_{c,s}=\delta\Omega_{c,s}\dd t$ and $\tau$ is the lag relaxation time. We take constant component moments of inertia and angular momenta $L_a=I_a\Omega_a$. We neglect the dependence of one component's momentum on the other's velocity due, or entrainment \citep{Prix:2001xc, Prix:2002jn}.
`Singular' dynamics use independent Wiener increments, $\dd\eta_a=\sqrt{Q_a}\dd W_a$, with $\langle\dd W_a\dd W_b\rangle=\delta_{ab}\dd t$ \citep{Meyers:2021myb}. The dynamics in the singular model has been teased out in the analytical solutions in both the time- and frequency-domains \citep{Meyers:2021myb, Meyers:2021slh, Bernardo:2026opv, ONeill:2024uiw}. `Minimal' dynamics set $x_c\dd\eta_c+x_s\dd\eta_s=0$, with $\dd\eta_c=\sqrt{A_{\min}/\tau}\dd W$ \citep{Antonelli:2022gqw}. At fixed total stochastic angular momentum and stationary initialization, the time-domain covariance of the crust angular velocity residuals can be written as
\begin{equation}
 C_{\Omega_c}(\Delta)=\frac{A_{\min}}2e^{-|\Delta|/\tau} \,.
 \label{eq:minimal-normalization}
\end{equation}
The integrated phase covariance is $A_{\min}H(u,v)/2$, where $u,v$ are times since initialization and \citep{Antonelli:2022gqw}
\begin{align}
 H(u,v)&=2\tau\min(u,v)\nonumber\\
 &\quad-\tau^2\big[1+e^{-|u-v|/\tau}-e^{-u/\tau}-e^{-v/\tau}\big] \, .
\end{align}

The measurement covariance includes scalable white noise,
\begin{equation}
 \sigma_{\phi,i,{\rm eff}}^2=F^2\sigma_{\phi,i}^2+(2\pi\nu_{{\rm nominal},i}J)^2 \,,
 \label{eq:white-noise}
\end{equation}
where $F$ is EFAC, which rescales the measurement uncertainty, and $J$ is EQUAD, an additional uncertainty added in quadrature \citep{2014MNRAS.437.3004L, 2016MNRAS.458.1267V, Goncharov:2024fsi}. The nominal secular spin frequency $\nu_{{\rm nominal},i}$ is the frequency used to convert timing uncertainties to phase uncertainties. All reported fits include $F$ and $J$, with normalized uniform priors on $\log_{10}F\in[-1,1]$ and $\log_{10}(J/{\rm s})\in[-8,-2]$. These parameters describe independent measurement variance in addition to the spin covariance.
For J1359-6038, the retained timing uncertainties already include the TEMPO2 calibration with EFAC $=1.18585$ and EQUAD $=17.03\,\mu{\rm s}$. The fitted $F$ and $J$ therefore act on these calibrated uncertainties. Vela's timing uncertainties have no prior rescaling.
For B1937+21, we use $\phi=2\pi F_0r_{\rm TOA}$ and $\sigma_\phi=2\pi F_0\sigma_{\rm TOA}$, where $r_{\rm TOA}$ and $\sigma_{\rm TOA}$ are the timing residual and its uncertainty, respectively, and $F_0$ is the rotational frequency.

For B1937+21, the HDF5 uncertainties are uncalibrated TOA errors. In phase units, we apply the published backend calibration,
\begin{equation}
 N_0=\operatorname{diag}\!\left[F_b^2(\sigma_{\phi,i}^2+J_{b,\phi}^2)\right]
       +\sum_{e,b}E_{b,\phi}^2u_{eb}u_{eb}^{T} \,,
 \label{eq:b1937-noise}
\end{equation}
where $F_b$ is the published EFAC, $J_b$ is T2EQUAD, and $E_b$ is ECORR \citep{2015ApJ...813...65N}. Here, $b$ labels the backend, $e$ labels the observing epoch, and $u_{eb}$ selects the channels recorded with backend $b$ during epoch $e$. ECORR is shared among those channels. T2EQUAD is added before multiplication by the backend EFAC. We apply this calibration once. The additional fitted $F$ and $J$ are then applied according to \eqref{eq:white-noise}.

Within each observing epoch, the phase measurements are combined using generalized least squares with weights $w=N_0^{-1}\boldsymbol{1}/(\boldsymbol{1}^{T}N_0^{-1}\boldsymbol{1})$, where $N_0$ is the calibrated white noise covariance \eqref{eq:b1937-noise}, and $\boldsymbol{1}$ is a vector of ones over the TOAs in that epoch. For additional data product in which phase and dispersion measure are jointly estimated within each epoch, we fit the two columns $[\boldsymbol{1},\boldsymbol{d}]$ using a scaled QR decomposition, preserving the covariance between the fitted phase and dispersion measure. The elements of $\boldsymbol{d}$ are $d_i=2\pi F_0K_{\rm disp}/f_{{\rm SSB},i}^2$, where $i$ labels the TOAs within the observing epoch, $F_0$ is the rotational frequency, and $f_{{\rm SSB},i}$ is the barycentric radio frequency in MHz. We use the standard dispersion constant $K_{\rm disp}$ \citep{2007MNRAS.378..493Y}. No temporal model for dispersion measure variations is fitted. All observing epochs are retained, including those with limited frequency separation.

For all three pulsars we use the Gaussian phase likelihood, with covariance $K$ including spin and measurement noise. The baseline timing design $X=[1,\Delta t,\Delta t^2/2]$ absorbs phase offset, frequency offset and spin down. With $\widehat\beta=(X^TK^{-1}X)^{-1}X^TK^{-1}y$ and $\boldsymbol{e}=y-X\widehat\beta$,
\begin{align}
 \log p(y\mid\theta,M)=-\frac12\big[&\boldsymbol{e}^TK^{-1}\boldsymbol{e}+\log|K|\nonumber\\
 &+\log|X^TK^{-1}X|\nonumber\\
 &+(N-p)\log(2\pi)\big] \, .
 \label{eq:likelihood}
\end{align}

Both models treat the linearized timing coefficients identically when they are marginalized. The state evolution is initialized 30 days before the first observing epoch. At this time, the singular residual states are set to zero, while the angular velocity in the minimal model is initialized from its stationary distribution. For Vela, we repeat the calculation with initialization times 10 and 100 days before the first epoch.

For Vela and B1937+21, the phase data and timing model are transformed using the model covariance before the timing coefficients are marginalized with a QR decomposition. Selected likelihood values and conditional moments are independently checked at high numerical precision, including limiting relaxation time cases. The principal evidence calculations are repeated with independent runs, more temperatures and longer chains, and are required to agree within their Monte Carlo uncertainties. The J1359-6038 evidence calculation follows the established phase analysis. Secondary Vela comparisons use one evidence run per fit, so their quoted uncertainties do not include variation between independent runs. For B1937+21, posterior sampling uses three independent ensembles and is required to satisfy effective sample size, chain length and standard convergence criteria. The full timing comparison retains all observing channels and 278 independent timing parameters, including 262 DMX coefficients. Numerically redundant timing directions are removed using a fixed criterion that is independent of the spin parameters. Evidence ratios are compared only between models using the same timing treatment.

For each reconstruction, we average the conditional phase predictions over posterior parameter samples. The uncertainty includes both the conditional variance and the variation of the conditional means, as well as uncertainty in the marginalized timing coefficients. The plotted bands show the posterior mean plus or minus one standard deviation. For display, the same quadratic projection is applied to the crust and superfluid reconstructions, with the covariance transformed consistently. This removes the fitted phase, frequency and spin down trends without introducing a separate fit to the superfluid prediction. For B1937+21, reconstruction stability is checked using independent ensembles and increasing posterior sample sizes. The crust reconstructions and the minimal superfluid reconstruction pass these checks. The singular superfluid variance does not, because a small number of allowed parameter values produce very large variances. Figure~\ref{fig:b1937} shows conditional superfluid means at specified superfluid moment of inertia fractions for both models. Each singular curve is obtained from a separate fit at the stated fraction, while the minimal curves use a single fit and scale as $-(1-x_s)/x_s$.

For the stochastic DM tests of J1359-6038 and Vela, we write the phase observations as
\begin{equation}
 y_i=X_i\beta+\delta\phi_{c,i}+d_i\delta{\rm DM}(t_i)+\epsilon_i,
 \qquad
 d_i=\frac{2\pi\nu_{{\rm export},i}K_{\rm disp}}{f_{{\rm SSB},i}^2} \,,
 \label{eq:dm-observation}
\end{equation}
where $\nu_{{\rm export},i}$ is the spin frequency used to convert the timing residuals to phase, $f_{{\rm SSB},i}$ is the barycentric radio frequency in MHz, and $K_{\rm disp}$ is the standard dispersion constant. We model $\delta{\rm DM}(t)$ as a Gaussian red process with temporal spectral index $\gamma=2$, $8/3$ or $4$. With $D={\rm diag}(d_i)$, the phase covariance becomes $K_{\rm spin}+DK_{\rm DM}D^T+N$. The lowest temporal frequency included in $K_{\rm DM}$ is $1/(10T)$, where $T$ is the observing span. The DM amplitudes are specified through the expected rms after removal of a linear trend at the observation epochs. The Gaussian DM coefficients are marginalized analytically in the likelihood.

For J1359-6038, later MeerTime observations give a linearly detrended DM rms of $0.013169\,{\rm pc\,cm^{-3}}$ over 4.02 years \citep{2024MNRAS.530.1581K}. Applying this amplitude to the earlier 3.46 year interval analyzed is a conditional assumption. For $\gamma=2$, $8/3$ and $4$, respectively, the comparisons give $\ln B_{s,m}=9.72\pm0.09$, $7.34\pm0.09$ and $6.59\pm0.09$. For Vela, previous observations indicate DM variability but do not determine its rms during the 307 day interval considered here \citep{2016MNRAS.459.3104S,2024PASA...41...80L}. We use $0.3\,{\rm pc\,cm^{-3}}$ as a stress test, obtaining $\ln B_{s,m}=6.80\pm0.13$, $12.53\pm0.07$ and $18.90\pm0.07$ for the same three temporal spectral indices. These amplitudes are imposed for the sensitivity tests and are not measurements of the DM variability in the analyzed intervals. For B1937+21, the wider radio frequency coverage allows DM offsets to be fitted within individual observing epochs without assuming a temporal spectrum or fixed stochastic DM amplitude, as described above. Evidence calculations use the same stepping stone procedure and numerical convergence checks described for the spin noise analysis.

\subsection{Goodness of fit}
\label{subsec:residual-adequacy}

One way to assess model fit is to check whether the residuals are uncorrelated and follow a standard normal distribution. We did that using the MAP values. However, to account for uncertainty in the model parameters, a more robust approach is to use posterior predictive checks.
This diagnostic compares the observed data to simulations drawn from the posterior predictive distribution, with smaller values indicating poorer agreement between the model and the data.
To compute $p_{\rm PPC}$, we first whiten the residuals. If $K=LL^T$ is the covariance and $X$ is the timing design, we compute $w=(I-QQ^T)L^{-1}y$, where $Q$ is an orthonormal basis for $L^{-1}X$. The standardized residuals are $z_i=w_i/[1-(QQ^T)_{ii}]^{1/2}$.

We assess the residuals using a simulation-calibrated posterior-predictive diagnostic. Timing-model removal induces correlations between the whitened residuals, which are retained in the simulation reference. For each fit, we draw 1,024 parameter vectors from the posterior, simulate observations at the measured epochs, and refit the spin and measurement-noise parameters using the original prior bounds and timing treatment. We examine residual scale, Gaussian shape, the largest absolute excursion, counts beyond three and four standard deviations, signed and squared temporal dependence, and variance across four equal-duration blocks. Temporal dependence uses eight logarithmic separation bins determined by the observing schedule, excluding bins with fewer than eight pairs.

The smallest upper-tail rank probability among these statistics is calibrated against the same minimum computed for the simulated
realizations. The resulting $p_{\rm PPC}$ accounts for inspecting several statistics and is reported in Table~\ref{tab:phase-results}.

\section{Power spectra and their interpretation}
\label{sec:power-spectra}

We give the explicit crust angular velocity and phase spectra; before removal of long term timing trends. The dynamics are given by (\ref{eq:eom_crust}-\ref{eq:eom_superfluid}). We denote by $\dot{\eta}_a=\dd\eta_a/\dd t$ the white stochastic torques.

We use the Fourier transform convention;
\begin{align}
\label{eq:fourier_transform} \widetilde u(\omega)
 &=\int_{-\infty}^{\infty}u(t)e^{+i\omega t}\dd t \,, \\
\label{eq:fourier_transform_inverse} u(t)
 &=\int_{-\infty}^{\infty}\frac{\dd\omega}{2\pi}
 \widetilde u(\omega)e^{-i\omega t} \,.
\end{align}
For a stationary process, the two sided spectral density is defined by
\begin{equation}
 \left\langle\widetilde u(\omega)
 \widetilde u^*(\omega')\right\rangle
 =2\pi\delta(\omega-\omega')P_u(\omega) \,.
 \label{eq:psd-convention}
\end{equation}
For the nonstationary processes below, $P_u$ denotes the formal spectrum obtained through the Fourier response at $\omega\ne0$.

Fourier transforming (\ref{eq:eom_crust}-\ref{eq:eom_superfluid}) give
\begin{equation}
 \widetilde{\delta\Omega}_c
 =
 \frac{(-i\omega+x_c/\tau)\widetilde{\dot{\eta}}_c
       +(x_s/\tau)\widetilde{\dot{\eta}}_s}
      {(-i\omega)(-i\omega+\tau^{-1})} \,.
 \label{eq:psd-response}
\end{equation}
For singular dynamics \citep{Meyers:2021myb},
\begin{equation}
 \left\langle\dot{\eta}_a(t)\dot{\eta}_b(t')\right\rangle
 =Q_a\delta_{ab}\delta(t-t') \,,
\end{equation}
and the crust angular velocity spectrum is
\begin{equation}
 P_{\Omega_c}^{(s)}(\omega)
 =
 \frac{Q_c(\omega^2\tau^2+x_c^2)+x_s^2Q_s}
      {\omega^2(1+\omega^2\tau^2)} \,.
 \label{eq:psd-singular-spin}
\end{equation}
This is the expression in Appendix A of \cite{ONeill:2024uiw}, with $\tau_c=\tau/x_s$, $\tau_s=\tau/x_c$, and $Q_a=\sigma_a^2/I_a^2$ in their notation.

Minimal dynamics \citep{Antonelli:2022gqw} impose
\begin{equation}
 \dot{\eta}_s=-\frac{x_c}{x_s}\dot{\eta}_c \,,
\end{equation}
so that the stochastic torques conserve total angular momentum. With
\begin{equation}
 \left\langle\dot{\eta}_c(t)\dot{\eta}_c(t')\right\rangle
 =\frac{A_{\min}}{\tau}\delta(t-t') \,,
\end{equation}
the crust obeys
\begin{equation}
 \dot{\delta\Omega}_c
 =-\frac{\delta\Omega_c}{\tau}+\dot{\eta}_c \,,
\end{equation}
and its spectrum is
\begin{equation}
 P_{\Omega_c}^{(m)}(\omega)
 =\frac{A_{\min}\tau}{1+\omega^2\tau^2} \,.
 \label{eq:psd-minimal-spin}
\end{equation}
This is the pure internal noise limit of \cite{Antonelli:2022gqw}.

Since $\dot{\delta\phi}_c=\delta\Omega_c$, integration contributes a factor $\omega^{-2}$ to the spectrum:
\begin{align}
 P_{\phi_c}^{(s)}(\omega)
 &=
 \frac{Q_c(\omega^2\tau^2+x_c^2)+x_s^2Q_s}
      {\omega^4(1+\omega^2\tau^2)} \,,
 \label{eq:psd-singular-phase}
 \\
 P_{\phi_c}^{(m)}(\omega)
 &=
 \frac{A_{\min}\tau}
      {\omega^2(1+\omega^2\tau^2)} \,.
 \label{eq:psd-minimal-phase}
\end{align}
For a constant reference angular velocity $\Omega_0=2\pi\nu_0$, the corresponding timing perturbation is $\delta t_c=\delta\phi_c/\Omega_0$, up to an overall sign convention. Its formal one sided spectral density per Hz is
\begin{equation}
 S_{\delta t_c}^{(j)}(f)
 =\frac{2}{\Omega_0^2}P_{\phi_c}^{(j)}(2\pi f) \,,
 \qquad f>0 \,,\quad j\in\{s,m\} \,,
 \label{eq:psd-timing-seconds}
\end{equation}
where $f$ is the temporal Fourier frequency.

At low frequencies compared to the crust-core lag relaxation time, $\omega\tau\ll1$, the minimal angular velocity spectrum is flat and its formal phase spectrum varies as $\omega^{-2}$. The singular angular velocity and phase spectra instead vary as $\omega^{-2}$ and $\omega^{-4}$, respectively, when the stochastic torques are nonzero. At high frequencies compared to the relaxation time, both phase spectra vary as $\omega^{-4}$ with nonzero crust stochastic torques. In the singular model, the superfluid torque contributes an additional term that varies as $\omega^{-6}$.

The singular angular velocity is nonstationary because independent stochastic torques allow the total angular momentum to diffuse \citep{Meyers:2021myb, Bernardo:2026opv}. Its covariance depends on both epochs rather than only on their separation, and neither the angular velocity nor its integrated phase has an ordinary stationary spectrum. Consequently, the formal expressions above cannot be inverted with the Wiener--Khinchin theorem to obtain the finite observation covariance.

For minimal dynamics, the angular velocity is a stationary process with covariance
\begin{equation}
 C_{\Omega_c}^{(m)}(\Delta)
 =\frac{A_{\min}}{2}e^{-|\Delta|/\tau}
 =\int_{-\infty}^{\infty}
   \frac{\dd\omega}{2\pi}
   P_{\Omega_c}^{(m)}(\omega)e^{-i\omega\Delta} \,.
 \label{eq:psd-minimal-wk}
\end{equation}
The phase obtained by integrating this process is nonstationary \citep{Antonelli:2022gqw}. Choosing $\delta\phi_c(t_0)=0$ and defining $u=t-t_0\geq0$ gives
\begin{equation}
 \operatorname{Var}[\delta\phi_c(t)]
 =A_{\min}\left[\tau u-\tau^2(1-e^{-u/\tau})\right] \,,
 \label{eq:psd-phase-variance}
\end{equation}
which grows linearly at long times \citep{Bernardo:2026opv}.


\end{document}